# Surface Charge Doping for Ion-Pairing Criticality in Confined Electrolytes

Na Shen[a], Yabei Wu[a,b], Wenqing Zhang[a,b,*]

a Department of Materials Science and Engineering, Southern University of Science and Technology, Shenzhen, 518055, China.

b Shenzhen Municipal Key-Lab for Advanced Quantum Materials and Devices, and Guangdong Provincial Key Lab for Computational Science and Materials Design, Southern University of Science and Technology, Shenzhen, 518055, China.

## ABSTRACT

Dielectric confinement strengthens Coulomb correlations in quasi-two-dimensional electrolytes and can promote Bjerrum pairing in charge-neutral slits. Here we use a generalized Debye-Huckel-Bjerrum theory to show that weak surface charge changes this picture by stoichiometrically doping the slit with mobile counterions. These counterions maintain a finite screening floor, decouple microscopic pairing from macroscopic ionicity, and shift association-driven criticality to lower temperatures. The critical-temperature suppression collapses onto a single scaled perturbation variable, revealing how surface charge and dielectric confinement jointly control charged nanofluidic slits. Brownian-dynamics tests further show that the same counterions are not always fully bulk-like diffusive: at low intrinsic salt density, explicit wall charge slows in-plane diffusion, whereas at higher intrinsic density the wall-induced diffusion penalty decreases and the mobile-counterion description becomes dynamically accurate. These results identify surface charge as a thermodynamic doping field that tunes both correlated ionic stability and the diffusion mechanism in nanofluidic confinement.

## I. INTRODUCTION

Angstrom-scale slits display ionic transport unlike that of bulk electrolytes, including nonlinear conductance and memory - like responses arising from strong Coulomb correlations due to dielectric mismatch confinement. [1–12]. In the neutral-surface limit, these correlations drive Bjerrum association in quasi-two-dimensional (quasi-2D) electrolytes and can approach a Kosterlitz-Thouless (KT)-like pairing scenario at low temperature or under strong confinement [3,11,13,14]. This picture provides an useful reference, but most nanofluidic interfaces carry surface charges and require compensating counterions inside the slit.

This raises the central question, whether the paired, low-ionicity state predicted by neutral-surface theories survives in a charged slit. This weak- charge limit is experimentally relevant, such as atomically flat graphite, hBN and $MoS_2$ angstrom-scale slits report surface charge densities of 20, 120 and 300 μC $m^{-2}$, respectively [1]. Although these values are small on macroscopic electrochemical scales, in an angstrom-scale slit they can correspond to a compensating counterion population comparable to, or larger than, the intrinsic free-ion population at low salt density. Electroneutrality therefore requires the surface charge to be compensated by mobile counterions within the confined fluid, converting the neutral pairing problem into a stoichiometrically doped quasi-2D Coulomb fluid. These surface-induced counterions create two competing effects. Locally, they increase the majority-ion population and can enhance the probability that minority ions find oppositely charged partners, thereby favouring Bjerrum association. Collectively, however, the same counterions impose a residual screening floor that supresses the long-range Coulomb attraction needed for association-driven criticality. How this weak-surface-charge doping modifies the spinodal boundary, the critical temperature, and the equivalence between microscopic pairing and macroscopic ionicity remains unknown.

Here we treat surface charge as a thermodynamic doping field in a dielectric-confined Coulomb fluid. We combine Bjerrum mass action with electroneutrality in a Debye-Huckel-Bjerrum theory that tracks intrinsic salt ions and compensating counterions separately. This formulation separates the local association from the collective screening.

This separation yields the central result. In a charged slit, microscopic ion pairing and macroscopic ionicity are not equivalent. Minority ions can be incorporated into neutral Bjerrum pairs almost completely, while excess majority counterions remain mobile and impose a residual screening floor. Consequently, surface charge suppresses the association-driven thermodynamic instability. It narrows the spinodal region, lowers the critical temperature, and stabilizes the homogeneous screening-rich state. The critical-temperature suppression further collapses onto a scaled perturbation variable that combines surface charge with the dielectric-confinement length, providing a compact design rule for charged nanofluidic slits.

The same mechanism also shapes in-plane diffusion. At low intrinsic salt density, electroneutrality creates a counterion-rich carrier population whose density is set mainly by surface-charge compensation rather than salt dissociation. Explicit-wall Brownian-dynamics simulations show that these carriers remain mobile, but are not dynamically completely free. Charged wall reduce $D_{xy}$ relative to the mobile-counterion equivalent model. As $\rho_0/\sigma$ increases, intrinsic ions increasingly dominate the transport response, and the wall-induced slowdown becomes a secondary correction. Thus, charged interfaces distinguish a low-density wall-controlled counterion-diffusion regime from a higher-density salt-dominated regime.

This paper is organized as follows. Section II develops the thermodynamic framework and excess chemical potential for a surface-charge-doped quasi-2D electrolyte. Section III analyses the decoupling between pairing, ionicity and stability, derives the critical scaling law, and benchmarks the associated in-plane diffusion response using explicit-wall Brownian dynamics. Section IV summarizes how surface charge, dielectric contrast and slit geometry control correlated ionic states.

# II. THEORY

## 2.1 Electrostatic Interaction and Screening in a Quasi-2D Slit

We consider a monovalent 1:1 electrolyte confined in an angstrom-scale slit of height $h \simeq 0.7$ nm, Two parallel planar walls bound the slit, and each wall carries a uniform negative surface charge density $-\sigma/2$. So that $\sigma > 0$ denotes the total compensating charge density required in the confined fluid. The conserved intrinsic salt density per unit area is $\rho_0$, and the area density of neutral Bjerrum pairs is $\rho_2$. Electroneutrality requires an excess counterions density, $\rho_{\mathrm{s}} = \sigma/e$. Taking anions as the minority species, we denote the free anion density by $m = \rho_-^{free}$. Electroneutrality then fixes the free cation density as $\rho_+^{free} = m + \rho_{\mathrm{s}}$. The total mobile charge density entering the Debye response is

$$\rho_1 = \rho_+^{free} + \rho_-^{free} = 2m + \rho_{\mathrm{s}}. \quad (1).$$

We focus on the weak surface-coupling regime, $h/\lambda_{\mathrm{GC}} \ll 1$, where $\lambda_{\mathrm{GC}}$ is the length scale at which the electrostatic attraction between a charged wall and a counterion becomes comparable to $k_{\mathrm{B}}T$. When this length is much larger than the slit height, the surface field is too weak to trap the counterions. The counterions therefore remain mobile in the plane of the slit and must be treated as part of the same quasi-two-dimensional Coulomb fluid as the intrinsic salt ions.

To benchmark this assumption, we perform BD simulations under matched parameters. The slit height is set to 7 Å and the microscopic hard-core cutoff is matched to the theoretical value $r_0$ = 0.95 Å, The results are compared with the theoretical results in Sec. III.

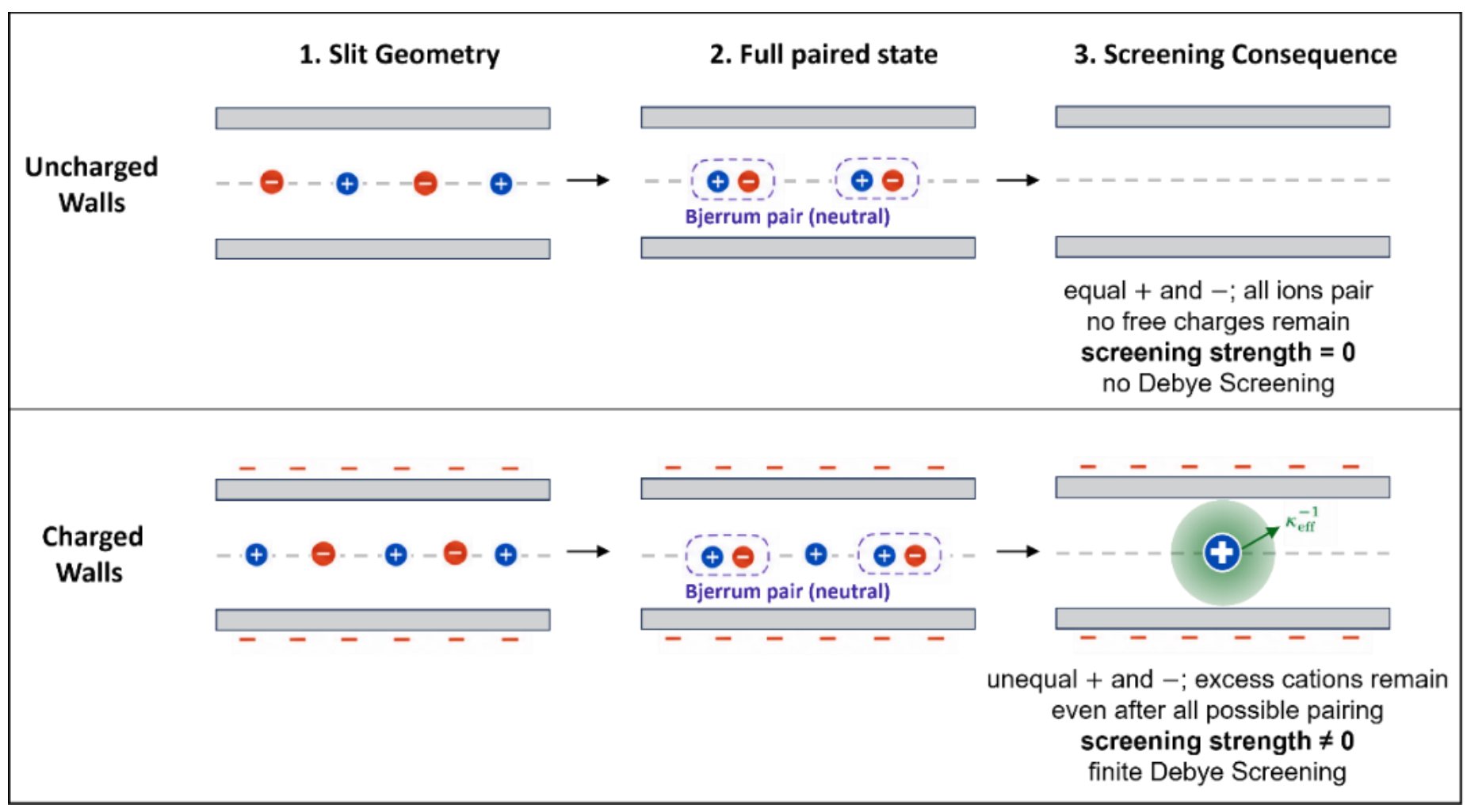

Figure 1. A neutral slit can approach a paired, carrier-depleted state, whereas a charged slit must contain mobile counterions, these counterions provide a finite screening background even when intrinsic ions bind into neutral Bjerrum pairs.

Using the reduced length $r^* = r/r_0$ and reduced dielectric-confinement length $\xi^* = \xi/r_0$, the dimensionless bare interaction between ions i and j can be written as [3,12,23]

$$\beta U_{ij}^{(0)}(r^*) = -\frac{z_i z_j}{T^*}\ln\left(\frac{r^*}{r^* + \xi^*}\right), \tag{2}$$

where $Z = \pm 1$ for a monovalent electrolyte, and

$$T^* = \frac{2\pi\epsilon_0\epsilon_\parallel h k_\mathrm{B}T}{Z^2 e^2}. \tag{3}$$

Here $\epsilon_\parallel$ is the effective dielectric constant inside the slit, $\epsilon_m$ is that of the surrounding surface medium, and the dielectric-confinement length is approximately $\xi \simeq \frac{\epsilon_\parallel h}{2\epsilon_m}$.

Linearizing the effective quasi-2D Poisson-Boltzmann equation around the homogeneous mobile-ion background gives effective screening strength

$$\kappa_\mathrm{eff}^2 = \frac{Z^2 e^2}{\epsilon_0\epsilon_\parallel h k_\mathrm{B}T}\rho_1 = \frac{2\pi}{T^*}\rho_1. \tag{4}$$

Equivalently, in reduced units,

$$\kappa_\mathrm{eff}^{*2} = \frac{2\pi}{T^*}\rho_1^*. \tag{5}$$

Equations (1), (4) and (5) contain the central consequence of surface-charge doping. In the neutral reference system $\sigma = 0$, Bjerrum association depletes the free intrinsic salt and vanishes the screening strength. However, for $\sigma > 0$, strong Bjerrum association still leaves $\rho_1$ to $\rho_s$, so $\kappa_\mathrm{eff}$ remains finite. This screening floor causes the decoupling between local Bjerrum association and macroscopic ionicity discussed below.

## 2.2 Thermodynamic Equilibrium with Broken Species Symmetry

We next formulate the thermodynamic description of the surface-charge-doped electrolyte. Surface charge breaks the equality of the two ionic populations. Taking anions as the minority species, we denote their free density by $m$. Electroneutrality then fixes $\rho_{-,\mathrm{free}} = m$ and $\rho_{+,free} = m + \rho_s$, where $\rho_s$ is the areal density of surface-induced counterions required to compensate the charged walls. Intrinsic salt conservation gives $\rho_0 = m + \rho_2$, where $\rho_2$ denoting the density of neutral Bjerrum pairs.

The Helmholtz free-energy density per unit area, in units of $k_\mathrm{B}T$, is [3]

$$\beta f = \beta f^\mathrm{el} + \sum_{i=\pm}\rho_i\,[\ln(\rho_i\Lambda^2) - 1] + \rho_2[\ln(\rho_2\Lambda^2) - 1] - \rho_2\ln\zeta_2\,. \tag{6}$$

Here $f^{\mathrm{el}}$ is the electrostatic excess free energy from Debye charging and depends on the total mobile free-charge density $\rho_1$. Neutral Bjerrum pairs contribute translational entropy and an internal configurational partition function $\zeta_2$. They do not enter Debye screening directly and affect screening only by depleting free ions.

The internal partition function of a bound pair is

$$\zeta_2^*(T^*) = 2\pi \int_1^{R_{\mathrm{B}}^*} r^* \log\left(\frac{r^* + \xi^*}{r^*}\right)^{1/T^*} dr^*, \tag{7}$$

where $R_{\mathrm{B}}^*$ is the reduced Bjerrum cutoff. This expression gives the configurational partition weight of bound oppositely charged pairs, defined by the condition that their attractive Coulomb energy exceeds the thermal energy scale $k_{\mathrm{B}}T$.

Chemical equilibrium imposes $\mu_2 = \mu_+ + \mu_-$. Using Eq. (6), we obtain the generalized mass-action law

$$\rho_2^* = \rho_{+,\mathrm{free}}{}^* \, \rho_{-,\mathrm{free}}{}^* \zeta_2^*(T^*) exp[2\mu_{\mathrm{ex}}^*(\rho_1^*, T^*)], \tag{8}$$

Where $\mu_{ex}^*$ is the reduced excess chemical potential per free ion.

$$\mu_{\mathrm{ex}}^* = \frac{\partial \beta f_{\mathrm{el}}^*}{\partial \rho_1^*}, \tag{9}$$

To quantify the macroscopic ionicity, we define the free fractions of each ionic species relative to the total density of each ionic species,

$$I_- = \frac{m}{\rho_0}, \qquad I_+ = \frac{m + \sigma}{\rho_0 + \sigma}. \tag{10}$$

The free fractions of all species is then

$$I = \frac{\rho_0 I_- + (\rho_0 + \sigma) I_+}{\rho_0 + (\rho_0 + \sigma)} = \frac{2m + \sigma}{2\rho_0 + \sigma}. \tag{11}$$

These definitions expose the separation introduced by surface charge. In the neutral system, strong association leads $m \to 0$ and derives the macroscopic ionicity ($I_{\pm}$, $I$) to zero. For $\rho_s > 0$, the same strong association gives $I_- \to 0$, while $I_+$ and $I$ remain finite because excess majority counterions cannot all be incorporated into neutral pairs. Minority ions can therefore be locally exhausted by Bjerrum association, whereas surface-induced counterions preserve a finite mobile-charge background. This stoichiometric constraint decouples microscopic pairing from macroscopic ionicity.

## 2.3 Stability Criterion and Critical Point

We next evaluate thermodynamic stability of the homogeneous phase, the spinodal corresponds to the loss of convexity of the constrained equilibrium free energy. We obtain the stability boundary by differentiating the mass-action branch at fixed $T^*$ and $\sigma^*$. From Eq. (8), the logarithmic slope of the neutral-pair density is

$$A(m,T^*;\sigma^*) \equiv \frac{\partial \ln\rho_2^*}{\partial m} = \frac{1}{m} + \frac{1}{m+\sigma^*} + 4\frac{\partial \mu_{\rm ex}^*}{\partial \rho_1^*}. \tag{12}$$

The first two terms arise from the ideal entropy of the two free ionic species, and the last term gives the electrostatic contribution to association. The spinodal is reached when the stabilizing entropic curvature is exactly compensated by the attractive electrostatic contribution. This condition corresponds to $A = 0$ and its derivative being zero,

$$A(m_c, T_c^*;\sigma^*) = 0, \tag{13}$$

and

$$\left(\frac{\partial A}{\partial \ln m}\right)_{m=m_c, T^*=T_c^*} = 0. \tag{14}$$

These equations expose a competition between association-induced carrier depletion and counterion-induced residual screening. Association removes free intrinsic ions and enhances the long-range Coulomb attraction that drives the instability, whereas surface charge injects excess counterions that maintain a finite screening floor, which stabilizes the homogeneous phase and depresses the critical temperature. We now use this framework to determine how this competition reshapes the pairing landscape, the stability boundary, and the resulting transport regimes.

# III. RESULTS AND DISCUSSION

## 3.1 Phase Diagram under Surface-Charge Doping

We first examine how surface charge reshapes the pairing landscape. In the neutral reference system, the density-weighted free ion fraction nearly vanishes at low $T^*$, consistent with strong Bjerrum association and symmetric depletion of cations and anions (Fig. 2a). With finite surface charge, this fully depleted limit is preempted. Even when the intrinsic density falls below $10^{-5}$ atom/Å$^2$, the free ion fraction remains finite (Fig. 2b), showing that surface-induced counterions impose a carrier floor absent from the neutral electrolyte.

A closer look at Fig. 2b separates this response into two regimes. At low intrinsic density, the free carriers are dominated by counterions required by electroneutrality, which produces a nearly $T^*$-insensitive conductive floor. At higher intrinsic salt density, the counterion excess becomes subdominant, and the free ion fraction recovers a nonlinear crossover governed by the intrinsic association equilibrium. Thus, surface charge does not simply shift the neutral pairing map, it changes which population controls the free charge.

The species-resolved free fractions $I_-$ and $I_+$ reveal the stoichiometric origin of this behavior. In the low-temperature strong-coupling regime, Bjerrum pairing nearly exhausts minority anions, giving $\rho_-^{free,*}\simeq 0$ and hence $I_-\to 0$ (Fig. 2c). Once this minority reservoir is depleted, excess majority cations have no anionic partners and remain dissociated, $\rho_+^{\mathrm{free},*} \simeq \sigma^*$. The majority-ion free fraction therefore approaches the surface-charge-controlled value, $I_+ \to \frac{\sigma^*}{\rho_0^* + \sigma^*}$ as shown in Fig. 2d. Complete depletion of one species therefore does not imply depletion of all free carriers.

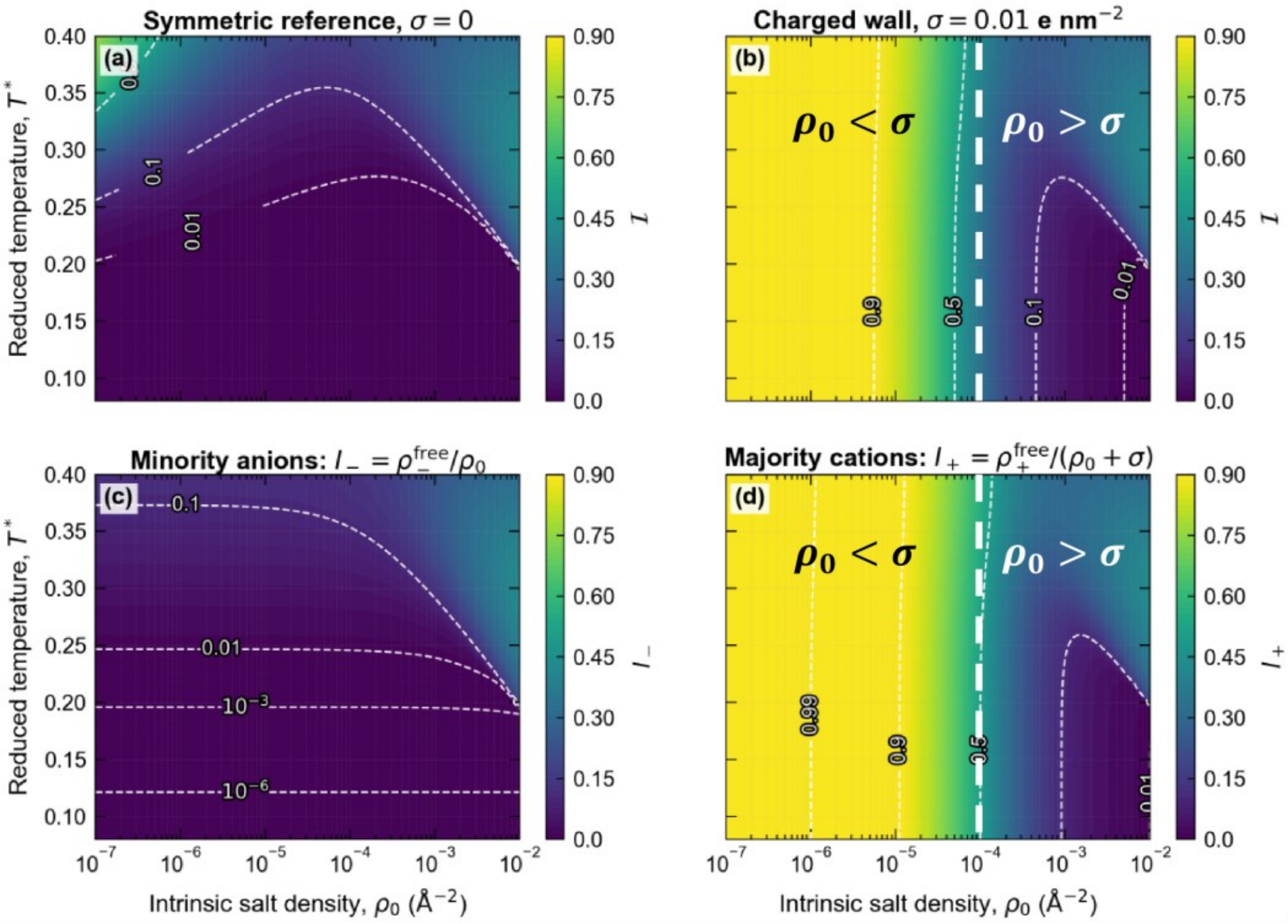

Figure 2. Stoichiometric decoupling of Bjerrum pairing and density-weighted ionicity. (a) Density-weighted ionicity $I$ for the neutral system without surface charge. (b) The same diagnostic for a charged slit, $\sigma = 10^{-2}\ e/nm^2$. (c) and (d) Species-resolved free fractions of minority intrinsic anions and majority cations, including counterions required by electroneutrality. At low $T^*$, minority anions are almost completely incorporated into neutral Bjerrum pairs, while a residual majority-cation density remains mobile.

We therefore use the BD results as a mechanistic cross-check of the surface-doped carrier floor. The BD results reproduce the key stoichiometric asymmetry predicted by the theory in Figure 3, minority anions are strongly depleted at low intrinsic density, whereas majority cations retain a finite free fraction set by surface-charge compensation. At higher density, the simulated free fractions are somewhat larger than the mean-field values, reflecting dynamical association criteria and many-body correlations beyond the analytic treatment. Overall, BD simulation can capture the theoretical transition trend well.

The low-temperature charged slit is therefore not a fully insulating paired fluid. It is better viewed as a fluid of neutral Bjerrum dipoles embedded in a residual background of mobile majority counterions. Surface charge thereby decouples microscopic association from macroscopic ionicity and creates a surface-doped finite-carrier state with no counterpart in the charge-symmetric reference system.

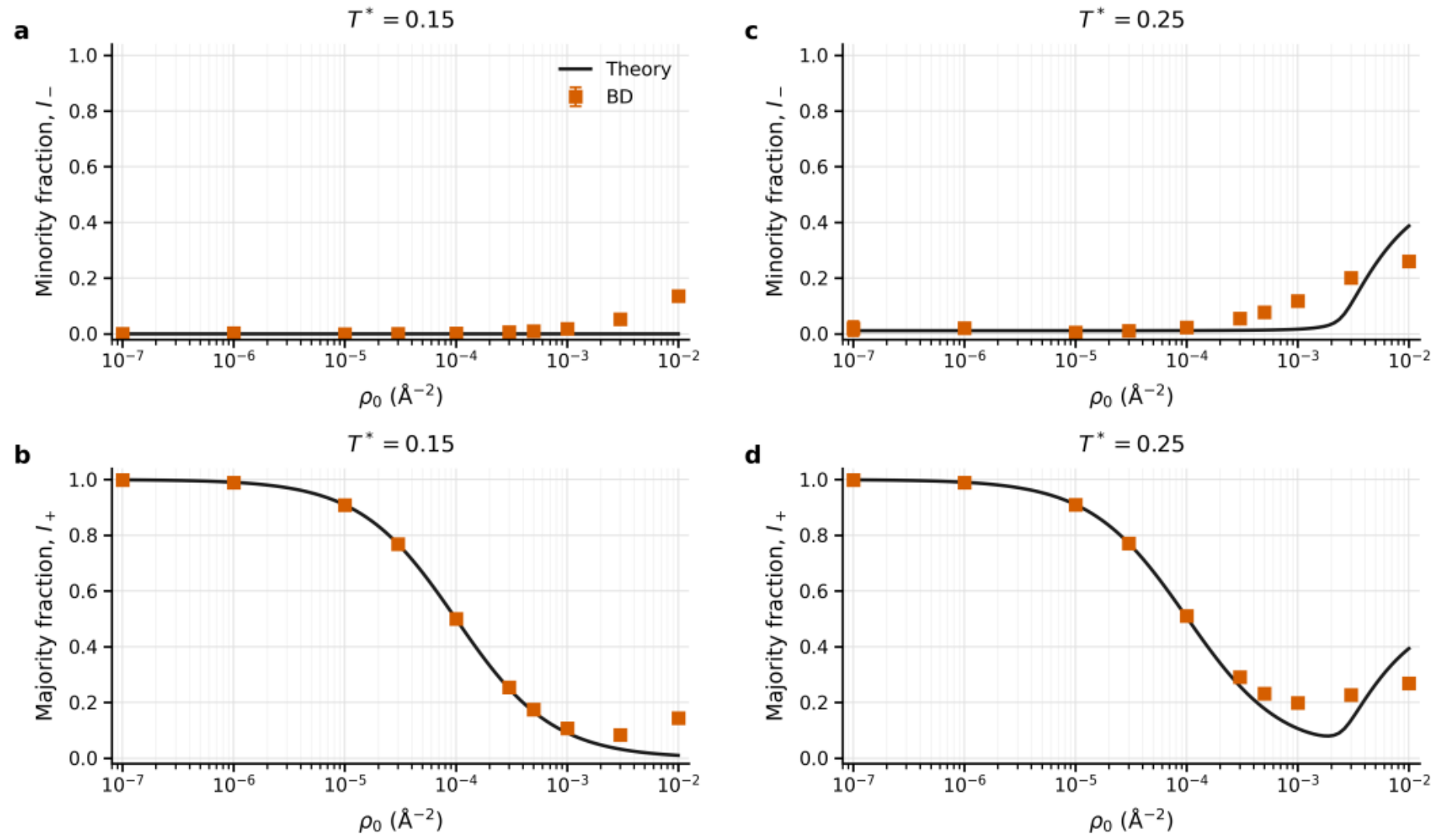

Figure 3. BD validation. Fixed-temperature line cuts of the species-resolved free fractions compared with explicit-wall Brownian dynamics simulations. Black curves are the theory used for Fig. 2, and orange squares are BD means from three independent seeds with error bars showing seed-to-seed standard deviations. The density scans capture the central prediction: minority anions (I-) are depleted in the strong-pairing, low-density regime, whereas majority cations (I+) retain a finite mobile-carrier fraction imposed by surface-charge compensation. The larger BD free fractions at high $\rho_0$ mark the residual quantitative uncertainty of the explicit-particle model rather than a failure of the surface-doping mechanism.

## 3.2 Dynamical consequence of explicit wall attraction

The equilibrium analysis above shows that surface charge creates a residual mobile-carrier population even when the intrinsic salt is strongly associated. We next ask whether these carriers are dynamically equivalent to those in the simplified mobile-counterion description, or whether explicit charged walls impose an additional mobility penalty. To this end, we compare Brownian dynamics simulations for three models: an explicit charged-wall slit, a mobile-counterion model without direct wall attraction, and a reference model in which wall attraction is removed. The in-plane diffusion coefficient $D_{xy}$ is extracted from the lateral mean-square displacement over the 100 ps time window.

Figure 4a,b shows the resulting $D_{xy}$ as a function of $\chi = \rho_0/\sigma$, which measures the intrinsic salt density relative to the compensating surface-charge density. At $T^* = 0.15$, the explicit charged-wall system displays a substantially smaller $D_{xy}$ than the two simplified models in the counterion-dominated regime, $\chi < 1$ (Fig. 4a). In this regime, most mobile carriers originate from electroneutrality rather than from intrinsic salt dissociation. Although these counterions remain free in the thermodynamic sense, their lateral motion is strongly hindered by direct wall attraction. Thus, the surface-induced carrier floor survives, but it is not dynamically equivalent to a freely mobile counterion reservoir.

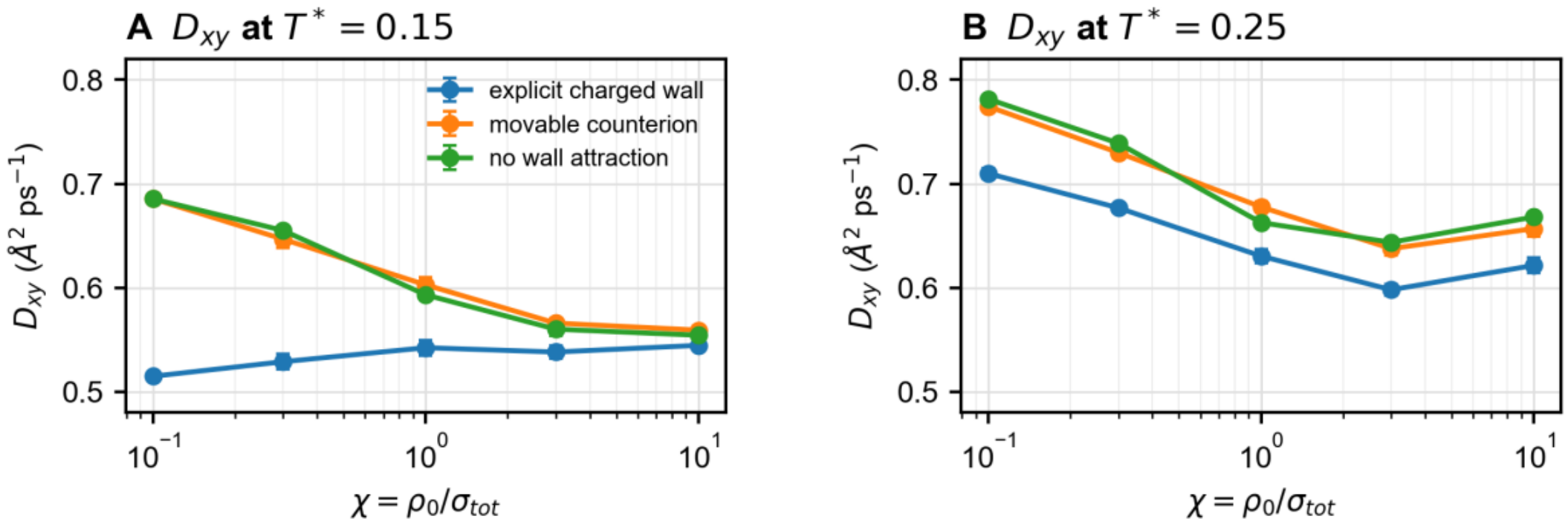

Figure 4. Explicit-wall BD validation. Complete boundary test at $\sigma = 1e-4\ e/Å^2$. (a,b) In-plane diffusivity $D_{xy}$ for explicit fixed wall charge, the mobile-counterion equivalent model, and the no-wall-attraction control at $T^*$= 0.15 and 0.25.

The comparison between the movable-counterion and no-wall-attraction models further identifies the origin of this slowdown. These two curves nearly overlap over the full range of $\chi$, whereas both lie above the explicit charged-wall result at low $\chi$. The dominant correction is therefore not the number of compensating ions itself, but the explicit ion-wall attraction that transiently traps or slows the counterions near the charged surfaces. This establishes a wall-controlled transport regime in which finite free ion does not imply bulk-like mobility.

The wall-induced mobility penalty weakens as $\chi$ increases. When the intrinsic salt density exceeds the compensating charge density, the dominant mobile population is no longer set solely by surface counterions. The explicit-wall curve then approaches the simplified-model curves, especially for $\chi \approx 3-10$. In this salt-dominated regime, direct wall attraction becomes a perturbative correction, and the mobile-counterion model provides a reasonable dynamical approximation to the explicit charged-wall system.

The same crossover is observed at the higher temperature ($T^* = 0.25$) (Fig. 4b), but with a reduced separation between the three models. Thermal fluctuations weaken the effect of wall attraction, so the explicit-wall diffusivity remains closer to the simplified-model values across the entire density range. The temperature dependence confirms that the slowdown at low $\chi$ is a dynamical consequence of wall-mediated residence and exchange rather than a loss of the thermodynamic carrier floor.

These results refine the surface-doped picture developed above. Surface charge guarantees a finite density of mobile charge carriers through electroneutrality, but the mobility of those carriers depends on whether transport is counterion-dominated or salt-dominated. At low $\chi$, explicit wall attraction converts the residual carrier floor into a wall-controlled transport state. At larger $\chi$, intrinsic ions dilute the wall-controlled counterion population, and the simplified mobile-counterion description becomes dynamically accurate to good approximation. Thus, surface charge decouples microscopic pairing from macroscopic ionicity, while explicit wall attraction controls how efficiently this residual ionicity contributes to lateral transport.

### 3.3 Suppression of Criticality

Having shown that surface charge decouples microscopic pairing from macroscopic ionicity, we next ask how this decoupling affects thermodynamic stability. In the neutral slit, strong Bjerrum association depletes free ions and can surpress the long-range Coulomb attraction that drives the homogeneous phase unstable. In a charged slit, this depletion competes with the counterion background imposed by electroneutrality.

The critical point provides a direct measure of this competition. For a representative charged slit, the pairing-equilibrium critical temperature decreases monotonically with increasing surface charge (Fig. 5a). Over $0 \leq \sigma \leq 0.05\ e/nm^2$, $T_c^*$ decreases from 0.242 in the neutral surface limit to 0.195 at $\sigma = 0.05\ e/nm^2$. The same stabilization appears in the spinodal curvature, where the negative-curvature window narrows as surface charge increases (Fig. 5c). Thus, surface charge weakens the association-driven instability rather than merely shifting the apparent salt concentration.

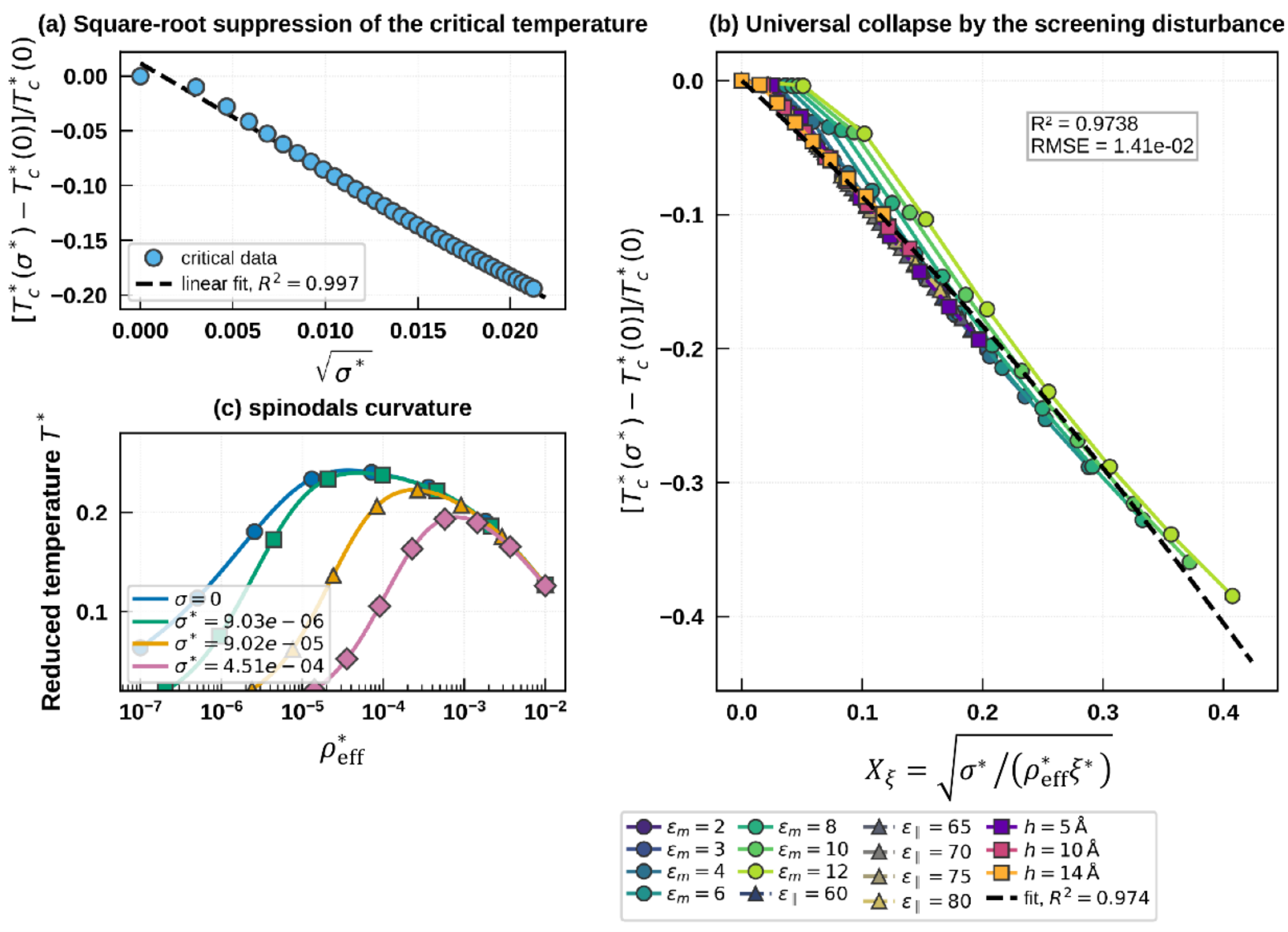

Figure 5. Surface charge suppresses pairing criticality. (a) Critical temperature as a function of surface-charge density. (b) Collapse of critical-temperature suppression against the scaled perturbation $X_\xi$. (c) Spinodal curvature for increasing surface-charge density, showing a narrowing unstable region shifted toward lower $T^*$ and higher intrinsic density.

This reduction of $T_c^*$ has a direct operational meaning. At the same temperature and intrinsic salt density, a charged slit lies farther from the phase separation curve than its neutral counterpart. Equivalently, reaching phase separation for homogeneous requires stronger confinement, lower temperature, or a larger intrinsic density once surface-charge compensation is present.

Fig. 5a establishes the suppression trend for a fixed confinement geometry. To test whether this trend is geometry-specific or controlled by a common perturbation scale, we vary the dielectric contrast and slit height. Surface charge provides the doping perturbation $\sigma^*$, $\rho_{1,c}^*$ sets the intrinsic screening scale of the neutral reference system, and the dielectric-confinement length $\xi^*$ sets the range over which this perturbation acts. These ingredients define

$$X_\xi = \sqrt{\frac{\sigma^*}{\rho_{1,c}^*(0)\xi^*}}. \tag{15}$$

When the critical-temperature suppression is plotted against $X_\xi$, data obtained by varying external dielectric constant and slit height collapse onto a single master curve (Fig. 5b). The collapse shows that surface charge, confinement geometry, and dielectric contrast enter through a combined screening perturbation rather than through independent controls. It also provides a practical design rule, since systems with the same $X_\xi$ exhibit comparable suppression of $T_c^*$.

Together, Fig. 5 shows that charged confinement reconstructs the stability landscape in two steps. Surface charge first imposes a residual mobile-charge background, which weakens the negative electrostatic curvature of the paired electrolyte. The resulting suppression is then organized by the scaled perturbation $X_\xi$, linking surface charge, dielectric confinement, and slit geometry in a single stability criterion.

# IV. CONCLUSION

We have shown that surface charge acts as a thermodynamic control field for correlated quasi-2D electrolytes. By extending the Debye-Huckel-Bjerrum framework to include mobile counterions required by electroneutrality, we find that charged confinement does more than shift the phase diagram of a neutral electrolyte. It reconstructs the relation between local ion association, collective screening, and macroscopic stability.

The central consequence is a stoichiometric decoupling between microscopic pairing and macroscopic ionicity. In a neutral slit, strong Bjerrum association can deplete both ionic species and produce a nearly total paired state. In a charged slit, surface-induced majority counterions exceed the pairing capacity of minority ions, minority ions can incorporated into neutral Bjerrum pairs almost completely, while majority counterions remain mobile. This residual mobile-charge background imposes a finite screening floor, narrows the negative-curvature region of the free energy, shifts the spinodal boundary to lower $T^*$, and reduces the critical temperature. The suppression collapses onto $X_\xi$, indicating that criticality is controlled by the combined surface-charge and dielectric-confinement scale.

Surface charge therefore divides the diffusion response into a low-density wall-associated counterion regime and a higher-density regime in which intrinsic-electrolyte diffusion is recovered. These results identify charged interfaces as active thermodynamic components of confined Coulomb fluids and suggest a route to tune correlated ionic states through surface charge density, dielectric contrast, and slit geometry.

## ACKNOWLEDGEMENTS

This work was supported in part by the National Natural Science Foundation of China (No. 92463310), the Guangdong Provincial Key Laboratory of Computational Science and Material Design (2019B030301001), and high-level special funds (G03050K002). W. Zhang also acknowledges support from the Guangdong Innovation Research Team Project (2017ZT07C062). We acknowledge computational support from the Center for Computational Science and Engineering at Southern University of Science and Technology.